\documentclass[epj,twocolumn]{webofc}
\usepackage[varg]{txfonts}
\usepackage{epstopdf}

\woctitle{Searching for the sources of galactic cosmic rays 2015}

\newcommand{\on}{N_{\mathrm{on}}}
\newcommand{\off}{N_{\mathrm{off}}}
\newcommand{\LM}{S_{\mathrm{LM}}}
\newcommand{\psr}{PSR~B1259--63/LS 2883}

\begin{document}
\captionsenglish

\title{Variability of VHE $\gamma$--ray emission from the binary PSR~B1259--63/LS 2883}

\author{Stanislav Stefanik\inst{1}\fnsep\thanks{\email{stefanik@ipnp.troja.mff.cuni.cz}} \and
        Dalibor Nosek\inst{1}
}

\institute{Institute of Particle and Nuclear Physics, Faculty of Mathematics and Physics, Charles University \\
V~Holesovickach~2, 180 00 Prague 8, Czech Republic
}

\abstract{
We examine changes of the $\gamma$-ray intensity observed from the direction of the binary system \mbox{\psr} during campaigns around its three periastron passages.
A simple and straightforward method is applied to the published data obtained with the 
Imaging Atmospheric Cherenkov Technique.
Regardless of many issues of the detection process, the method works only with numbers of very high energetic photons registered in the specified regions.
Within the realm of this scheme, we recognized changes attributable to the variations of the intrinsic source activity at high levels of significance.
}

\maketitle


\section{Introduction}
\label{intro}
\psr~is one of few $\gamma$--ray binaries detected in the very high energetic (VHE) regime.
It is known to consist out of a Be star with an equatorial disk tilted relative to the movement of the pulsar which orbits around the star with a period $T = 1237~\mathrm{days}$~\cite{Johnston}.
The variability of the source emission has been clearly observed in multiple wavebands from radio to VHE with the most prominent changes occurring around the periastron passages~\cite{hess_2013}.
The cause of this variations is believed to be the interaction of the pulsar with the stellar disk.
Significant variability of the VHE $\gamma$--ray flux around periastron~\cite{hess_2005, hess_2009} was observed by the H.E.S.S.~telescope array.
Although a considerable overlap of the pre-- and post--periastron data has not been obtained so far, a stronger evidence for the repetitive behaviour stemming from the orbital motion was claimed based on the correspondence of the newest data with the older lighcurve~\cite{hess_2013}.
Recently, detection of a GeV flare by the \emph{Fermi}--LAT~\cite{Abdo, Tam} and a lack of evidence for an attributable VHE flare~\cite{hess_2013} lead to the conclusion that the high energy (HE) photons have a different origin than the VHE ones.
In this paper, we reexamine the changes of VHE activity of \psr~during campaigns targeted at periods of periastron passages.
\par 
We deal with publicly available VHE data gathered by Cherenkov telescopes in the on/off observation mode~\cite{Berge}.
In this approach, the numbers of detected events $\on$ and $\off$ are extracted from the observed signal (on) and background (off) regions, respectively.
They are related by the on--off parameter $\alpha$ which is the ratio of exposures of signal and background regions.
Traditionally, the level of significance of the source presence in the on--source region is determined using the Li--Ma method~\cite{LiMa}.
In this work, we use its modification for the investigation of changes in the source intensity (see \cite{Nosek, Stefanik} for details).
A source parameter $\beta > 0$ is introduced to quantify the signal in the on--source region when compared to the background ascertained from the reference regions.
The core of the modified on--off method is the statement of the null hypothesis that a previously identified source of interest attains intensity of chosen value~$\beta$, i.e.~$\on = \alpha \beta \off$~\cite{Nosek}.
A level of significance for the rejection of the assumption of constant source activity is given by~\cite{Nosek, Stefanik}

\begin{equation}
\label{eq:lima}
\LM = s\sqrt{2}\left\lbrace \on \ln{ X_{1} } + \off \ln{ X_{2} } \right\rbrace^{\frac{1}{2}},
\end{equation}

\noindent
for the asymptotic Li--Ma statistics.
Here, the logarithmic arguments are $X_{1} = \frac{1+\alpha\beta}{\alpha\beta} \frac{\on}{\on + \off}$ and $X_{2} = (1+\alpha\beta) \frac{\off}{\on + \off} $.
The $s$--term in Eq.~(\ref{eq:lima}) is either $+1$ or $-1$ for observations of an excess ($\LM>0$) or deficit ($\LM<0$) of events, respectively.
\par
Li--Ma significance given in Eq.~(\ref{eq:lima}) asymptotically follows the normal distribution with zero mean and unit variance \cite{Nosek}.
Therefore, any inconsistency between the sample distributions of $\LM$ and the reference standard Gaussian distribution should be regarded as a sign of change in the tested $\gamma$--ray intensity.
It is also worth noting, that the test of the null hypothesis can be advantageously inverted to derive confidence intervals for the source parameter $\beta$ at a given level of significance.
Sequence of such confidence intervals then carries information about changes of the observed $\gamma$--ray activity.
\par 
The correct treatment of Poisson observables is inherent in the modified on--off method which straightforwardly exploits just the numbers of detected photons.
Its other benefit is that systematic uncertainties are fully accounted for and are already included in the confidence intervals for the source parameter.
The application of the method is thus very simple yet the range of possible utilizations and the interpretations of results remain utmost general.


\section{Data analysis}
\label{sec:analysis}

Data sets used in our study comprise results of observations of \psr~by the H.E.S.S.~telescope array during periastron passages in 2004~\cite{hess_2005}, 2007~\cite{hess_2009} and 2011~\cite{hess_2013}.
Presented numbers of on-- and off--source counts together with values of the parameter $\alpha$ were used as input.
Measured counts in each studied observational period ranged between over one hundred to several thousands.
Data were taken in configurations with different numbers of active telescopes.
\par
\begin{figure}
	\centering
		\includegraphics[width=\columnwidth]{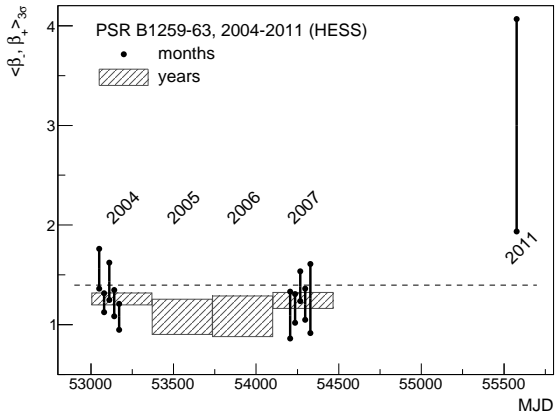}
		\caption{
			MJD--ordered sequence of confidence intervals at a $3\sigma$ level for the source parameter $\beta$ obtained for the $\gamma$--ray binary \psr~\cite{hess_2005, hess_2009, hess_2013}.
			Segments with points represent monthly intervals, hatched bands denote yearly intervals.
			The horizontal dashed line indicates the average value of the source parameter taken over all measurements.
			Interval for 2011 observations was estimated by accounting for the energy spectrum presented by the H.E.S.S.~collaboration, for details see Tab.~\ref{tab:CI} and corresponding text.
		}
		\label{fig:CI}
\end{figure}

We examined changes of the observed source intensity deduced from the H.E.S.S.~data by the means of confidence intervals for the source parameter $\beta$.
These intervals were constructed at a $3\sigma$ ($\approx 99.7\%$) level of significance for the sequence of monthly triplets $(\on,~\off,~\alpha)$ using Eq.~(\ref{eq:lima}) such that $|\LM(\on,~\off,~\alpha;~\beta)|<3$.
Furthermore, we combined the monthly data into larger sets according to calendar years and calculated the corresponding confidence intervals.
The results are visualized in Fig.~\ref{fig:CI} where the intervals are depicted as MJD--ordered time series.
Several non--overlapping monthly confidence intervals (black segments) indicate variations of the intensity observed from the source.
Annual observations provide intervals (hatched bands) which are consistent altogether.
Confidence interval for 2011 campaign depicted in Fig.~\ref{fig:CI} was obtained with the modified data set (see Tab.~\ref{tab:CI} and Section~\ref{sec:discussion}).

\begin{figure}
	\centering
		\includegraphics[width=\columnwidth]{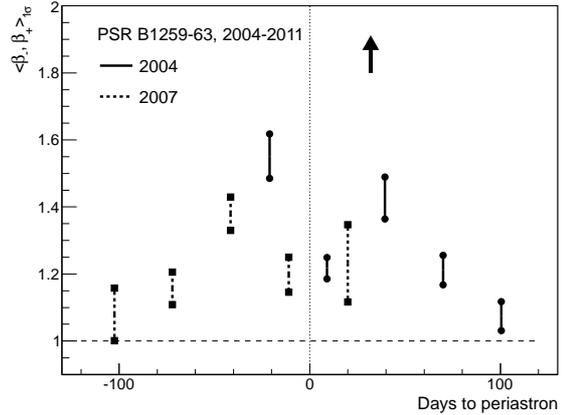}
		\caption{
			Confidence intervals at a $1\sigma$ level as a function of observation date with respect to the periastron occurrence which is marked by vertical dotted line.
			The horizontal dashed line denotes the reference value corresponding to the background--only dominated on--source region.
			Observations in 2011, $\langle \beta_{-},~\beta_{+} \rangle = \langle 3.57,~4.44 \rangle$, are indicated by the arrow.
			For further details see caption to Fig.~\ref{fig:CI}.
		}
		\label{fig:CI_phase}
\end{figure}

\par 
The observed data can be further depicted in a plot for confidence intervals at a $1\sigma$ level arranged according to the proximity of observations to the periastron date.
In Fig.~\ref{fig:CI_phase}, indications of a double--peak structure can be recognized with a local minimum around the date of periastron.
\par 
In order to quantify the trends of the source activity emerging in the sequence of confidence intervals we also performed test with the 2004 and 2007 data for the assumption of constant intensity.
The source activity to be tested was chosen as an average value of the ratio of observed and expected on--source events over the observational periods in 2007, i.e.~$\beta_{07} = \langle \on / \alpha \off \rangle = 1.21$.
Our results of testing changes of the source activity are presented as a quantile--quantile (QQ) plot.
For this purpose, Li--Ma significances were calculated according to Eq.~(\ref{eq:lima}) and then assigned to the quantiles of the standard normal distribution $N(0,1)$.
The quantiles were chosen to be \mbox{$k / (m+1)$} where $m$ denotes the number of observed events and $k = 1 \dots m$.
The observational time sequence is indicated as increasing sizes of markers.

\begin{figure}
	\centering
	\sidecaption
		\includegraphics[width=\columnwidth]{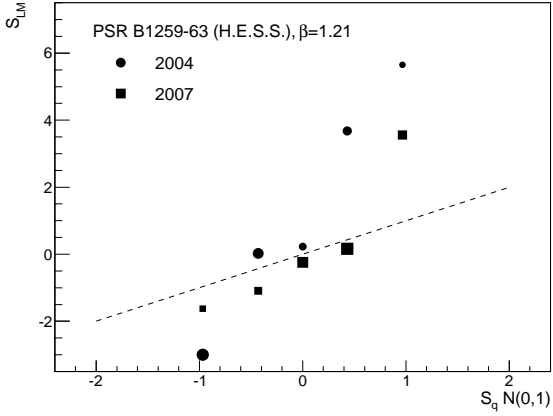}
		\caption{
			QQ--plot of asymptotic Li--Ma significances for the $\gamma$--ray events detected from the direction of \psr~\cite{hess_2005, hess_2009}.
			Average source parameter $\beta_{07} = 1.21$ calculated from 2007 measurements was tested.
			The chronological sequence of observations is given by increasing sizes of markers.
			The dashed line denotes the reference $45^{\circ}$ line through the origin.
		}
		\label{fig:QQ}
\end{figure}

In Fig.~\ref{fig:QQ}, a QQ--plot of the distribution of Li--Ma significances is shown.
The dashed line denotes the reference $45^{\circ}$ line through the origin on which the data points lie provided the hypothesis of constant source intensity of given value is true.
Most of the measurements conducted in 2007 (squares) are approximately consistent with the reference line apart from the minor downward shift which indicates that the value of intensity during these periods was slightly lower than the chosen one.
The only exceptional result in 2007 is obtained for the June data set which yields significances for the rejection of a steady source activity assumption $\LM > 3.6$.
On contrary, the 2004 sample significances (circles) exhibit considerable dispersion from the reference line.
Different skewness of the distributions of sample significances compared to the reference standard normal distribution hints that the source intensity changed.
At two occasions a significant excess of source activity is recognized with the significances $\LM > 3.6$.
In one case a considerable deficit of events is observed with $\LM \approx -3$.


\section{Discussion}
\label{sec:discussion}

\begin{table}
	\centering
	\caption{
	  	Numbers of detected on-- and off--source events ($E>500$~GeV) and the on--off parameter $\alpha$ for the H.E.S.S.~observations of \psr~during 2011~\cite{hess_2013}.
	  	$\langle \beta_{-},~\beta_{+} \rangle_{3\sigma}$ are confidence intervals for the source parameter $\beta$ constructed at a $3\sigma$ level.
	  	Asterisk denotes the modified data set.
	  }
  	\label{tab:CI}
	  \begin{tabular}{l c c c c}
		\hline
		Period & $\on$ & $\off$ & $\alpha$ & $\langle \beta_{-},~\beta_{+} \rangle_{3\sigma}$ \\
		\hline
		2011   & 112   & 365    & 0.077   & $\langle 2.85,~5.46 \rangle$   \\	
		2011*  & 80    & 365    & 0.077   & $\langle 1.93,~4.07 \rangle$   \\	
		\hline
	  \end{tabular}
\end{table}

The H.E.S.S.~collaboration stated that there is evidence for variability of integral flux above the energy threshold 380~GeV from \psr~in 2004~\cite{hess_2005} and also above 620~GeV in 2007~\cite{hess_2009}.
Moreover, the overall flux level was found to correspond to the one measured in 2004.
Observations in 2005 and 2006, i.e.~during the epoch between two consecutive periastron passages, did not provide significant detection of the source thus indicating its variable character~\cite{hess_2009}.
No variability on the time scale of individual nights during 2011 was proved~\cite{hess_2013}.
It was further stated that the 2011 flux above the energy threshold 1~TeV is consistent with the flux levels obtained in 2004 and 2007.
\par 
In the framework of the modified on--off method, we arrive at similar results with regard to the evidence for variability of observed $\gamma$--ray intensity in 2004 and 2007, see Figs.~\ref{fig:CI} and \ref{fig:QQ}.
Likewise, increase of the source emission before and after the periastron visible in Fig.~\ref{fig:CI_phase} agrees with the results of the lightcurve analysis~\cite{hess_2009, hess_2013}.
Interestingly, our results suggest that the source activity observed in 2011 was not consistent with the overall intensity in previous campaigns.
First row in Tab.~\ref{tab:CI} shows that the $99.7\%$ confidence interval for 2011 measurements is lying considerably higher than any other one deduced from earlier observations, compare with Fig.~\ref{fig:CI}.
This translates also in the significance $\LM = 9.6$ with which one can reject the assumption of the constant source intensity.
Here, the source parameter $\beta_{47} = 1.25$ to be tested was chosen as an average value taken from the 2004 and 2007 campaigns.
\par 
These results are not contradictory, however.
Firstly, due to the smaller data set the results of the 2011 measurements are less constraining than the previous ones~\cite{hess_2013}.
Secondly, we worked also with events with energies below 1~TeV, the threshold above which H.E.S.S.~determined the integral flux~\cite{hess_2013}.
Finally, it is worth noting that the H.E.S.S.~experiment measured considerably increased flux of photons at energies around 500~GeV, see Fig.1 in \cite{hess_2013}.
\par 
In an attempt to explain the inconsistencies we changed the number of photons observed in the first interval of the energy spectrum.
To this end, we assumed the same simple power law spectral shape with the index $\Gamma = 2.92$ as reported in the H.E.S.S.~analysis~\cite{hess_2013}.
The modified number of photons in the lowest energy bin was set as consistent with the power law derived from higher energy bins.
The resulting 2011 confidence interval for the source parameter $\beta$ stemming from this modification is given in the second row in~Tab.~\ref{tab:CI}.
This interval still indicates that the observed source activity above the energy threshold of roughly 500~GeV was in 2011 above the intensities deduced from previous campaigns provided the background remained stable throughout all observation periods.
\par 
In view of the 2011 detection of a HE $\gamma$--ray flare by the \emph{Fermi}--LAT~\cite{Abdo, Tam}, it may be suggestive to relate this flare to the higher level of the VHE activity as seen in our analysis.
However, when testing the `Pre--flare' and `Flare' data sets of H.E.S.S.~\cite{hess_2013} for deviations of the source intensity from its average value, the test significances for both periods are $|\LM|<1$.
Thus, we cannot reject the assumption of constant intensity during 2011 based on these data and, accordingly, the connection with the GeV flare is questionable.
\par 
The increase of the observed VHE activity may be attributable to the truly intrinsically variable emission or it may be some effect of the detection process.
We cannot exclude the latter possibility without further details on the used data, particularly on the exact detection thresholds in each observational campaign.
Nevertheless, we have no firm indication of inconsistency of the source intensity above 1~TeV in 2011 when compared to the previous observations.
Hence, we comply with the H.E.S.S.~conclusions about the overall agreement of integral fluxes above this energy.


\section{Conclusions}

We used the modified on--off method to judge whether significant changes of the $\gamma$--ray intensity from the direction of binary \psr~were observed during its three periastron passages.
Within our method, the problem of variable source activity can be treated using solely the numbers of detected events.
\par 
Temporal evolution of the observed source intensity is clearly recognized on the account of the modified on--off technique.
The results of our analysis of the data gathered by the H.E.S.S.~experiment clearly show that changes of the source activity occurred during 2004 and 2007 between individual months while the overall intensities in both seasons stayed in agreement.
Our findings on the data obtained above approximately 500~GeV in 2011 show that the observed source intensity was enhanced in this season with respect to earlier measurements.


\begin{acknowledgement}
This work was supported by the Czech Science Foundation grant 14-17501S.
\end{acknowledgement}


\end{document}